\documentclass[a4paper,12pt]{article}
\usepackage{amsfonts}
\usepackage{latexsym}
\usepackage{amssymb}
\usepackage{color}
\date{}

\begin{document}

            \newcommand{\be}{\begin{equation}}
            \newcommand{\beg}[1]{\begin{equation}\label{#1}}
            \newcommand{\ee}{\end{equation}\normalsize}
            \newcommand{\bee}[1]{\begin{equation}\label{#1}}
            \newcommand{\bey}{\begin{eqnarray}}
            \newcommand{\byy}[1]{\begin{eqnarray}\label{#1}}
            \newcommand{\eey}{\end{eqnarray}\normalsize}
            \newcommand{\beo}{\begin{eqnarray}\normalsize}
            \newcommand{\R}[1]{(\ref{#1})}
            \newcommand{\C}[1]{\cite{#1}}

            \newcommand{\mvec}[1]{\mbox{\boldmath{$#1$}}}
            \newcommand{\x}{(\!\mvec{x}, t)}
            \newcommand{\m}{\mvec{m}}
            \newcommand{\F}{{\cal F}}
            \newcommand{\n}{\mvec{n}}
            \newcommand{\argm}{(\m ,\mvec{x}, t)}
            \newcommand{\argn}{(\n ,\mvec{x}, t)}
            \newcommand{\T}[1]{\widetilde{#1}}
            \newcommand{\U}[1]{\underline{#1}}
            \newcommand{\V}[1]{\overline{#1}}
            \newcommand{\ub}[1]{\underbrace{#1}}
            \newcommand{\X}{\!\mvec{X} (\cdot)}
            \newcommand{\cd}{(\cdot)}
            \newcommand{\Q}{\mbox{\bf Q}}
            \newcommand{\p}{\partial_t}
            \newcommand{\z}{\!\mvec{z}}
            \newcommand{\bu}{\!\mvec{u}}
            \newcommand{\rr}{\!\mvec{r}}
            \newcommand{\w}{\!\mvec{w}}
            \newcommand{\g}{\!\mvec{g}}
            \newcommand{\D}{I\!\!D}
            \newcommand{\se}[1]{_{\mvec{;}#1}}
            \newcommand{\sek}[1]{_{\mvec{;}#1]}}            
            \newcommand{\seb}[1]{_{\mvec{;}#1)}}            
            \newcommand{\ko}[1]{_{\mvec{,}#1}}
            \newcommand{\ab}[1]{_{\mvec{|}#1}}
            \newcommand{\abb}[1]{_{\mvec{||}#1}}
            \newcommand{\td}{{^{\bullet}}}
            \newcommand{\eq}{{_{eq}}}
            \newcommand{\eqo}{{^{eq}}}
            \newcommand{\f}{\varphi}
            \newcommand{\rh}{\varrho}
            \newcommand{\dm}{\diamond\!}
            \newcommand{\seq}{\stackrel{_\bullet}{=}}
            \newcommand{\st}[2]{\stackrel{_#1}{#2}}
            \newcommand{\om}{\Omega}
            \newcommand{\emp}{\emptyset}
            \newcommand{\bt}{\bowtie}
            \newcommand{\btu}{\boxdot}
            \newcommand{\tup}{_\triangle}
            \newcommand{\tdo}{_\triangledown}
            \newcommand{\Ka}{\frac{\nu^A}{\Theta^A}}
            \newcommand{\K}[1]{\frac{1}{\Theta^{#1}}}
            \newcommand{\ap}{\approx}
            \newcommand{\bg}{\st{\Box}{=}}
            \newcommand{\si}{\simeq}
\newcommand{\Section}[1]{\section{\mbox{}\hspace{-.6cm}.\hspace{.4cm}#1}}
\newcommand{\Subsection}[1]{\subsection{\mbox{}\hspace{-.6cm}.\hspace{.4cm}
\em #1}}

\newcommand{\const}{\textit{const.}}
\newcommand{\vect}[1]{\underline{\ensuremath{#1}}}  
\newcommand{\abl}[2]{\ensuremath{\frac{\partial #1}{\partial #2}}}

\title{Constitutive Settings with regard to\\
Energy- and Entropy-Balances\\in Non-Equilibrium Thermodynamics:\\
the Thermodynamical Verification}
\author{W. Muschik\footnote{Corresponding author: muschik@physik.tu-berlin.de}
\thanks{In memory of Bogdan Maruszewski}
\\
Institut f\"ur Theoretische Physik\\
Technische Universit\"at Berlin\\
Hardenbergstr. 36\\D-10623 BERLIN,  Germany}
\maketitle
\abstract\noindent
Constitutive equations have to be in agreement with the energy- and entropy-balances.
For achieving that, the procedure of thermodynamical verification is introduced: 
Because heat flux and entropy flux as well as the time differentials of internal energy and
entropy are not independent of each other,  energy- and entropy-balances are connected with
each other by so-called internal settings laying down the theoretical frame of the applied
material description which is characterized by additional constitutive settings.
\vspace{.3cm}\newline\noindent
{\bf Hint:} This is a short-note: more details are can be found in the reference
below\footnote{see e.g.: Restuccia, L.; Jou, D.; Pavelka, M. On the identification of two internal
tensorial variables and a heat transport equation with internal, thermal viscosity and vorticity terms. {\em J. Non-Equilib. Thermodyn.} {\bf 2026}, {\em 51}, 1-18.} 
\vspace{.3cm}\newline\noindent
For elucidating the procedure of thermodynamical verification, three examples are considered
which use different settings.\\
\noindent\\
{\bf 1. A trivial example: Fourier heat conduction}\footnote{for a warm up}\\
Starting out with the energy- and entropy-balances \C{Chris}
\bee{a1}
\varrho\st{\bullet}{u}+\nabla\cdot\mvec{q}\ \st{1}{=}\ r,
\qquad \sigma\ \st{2}{=}\ \varrho\st{\bullet}{s}+\nabla\cdot\mvec{J},
\ee
(mass density $\varrho$, internal energy $u$, heat flux $\mvec{q}$, energy supply $r$,
entropy production $\sigma$, entropy $s$, entropy flux $\mvec{J}$, time derivative 
$\bullet $)  four different equal signs are introduced to better distinguish the steps of
thermodynamical verification:
\byy{a2}
\st{1}{=}\ \mbox{and}\ \st{2}{=} \mbox{for the balances,}\ &\st{c}{=}&\ 
\mbox{for a constitutive setting,}\\ \label{a3}
&\seq& \ \mbox{for an internal setting.}
\eey
The Fourier heat flux satisfies a constitutive equation, the entropy flux an internal setting
\bee{a4}
\mvec{q}\ \st{c}{=}\ -\kappa\nabla T,\qquad \mvec{J}\ \seq \ \mvec{q}/T
\ee
(heat conduction $\kappa$, temperature T).
Using \R{a1}$_2$ and \R{a4}$_2$ results in
\byy{a5}
\sigma-\varrho\st{\bullet}{s} &\st{2}{=}&\nabla\cdot(\mvec{q}/T)\ =\ \mvec{q}\cdot\nabla(1/T)
+(1/T)\nabla\cdot\mvec{q}\ \st{1}{=}\\ \label{a6} 
 &\st{1}{=}&\ \mvec{q}\cdot\nabla(1/T)+(1/T)(r-\varrho\st{\bullet}{u}).
\eey
Now the time differential of the entropy can be determined from \R{a6} by an internal setting
resulting in the entropy production 
\bee{a7}
-\varrho\st{\bullet}{s}\ \seq\ -\varrho(1/T)\st{\bullet}{u},\ \longrightarrow\ 
\sigma\ = \mvec{q}\cdot\nabla(1/T)+(1/T)r.
\ee
Inserting the constitutive setting \R{a4}$_1$ yields
\bee{a8}
\sigma \st{c}{=} -\kappa\nabla T\cdot\nabla(1/T)+(1/T)r\ =\ (\kappa /T^2)\nabla T\cdot\nabla T
+(1/T)r.
\ee

Because the entropy balance \R{a1}$_2$ is built up of three terms, two terms have to be chosen by
internal settings, here [$\st{\bullet}{s},\mvec{J}$] are chosen in \R{a7}$_1$ and \R{a4}$_2$.
The same result can be obtained by the internal setting [$\st{\bullet}{s},\sigma$] in \R{a7}
replacing $\mvec{J}$ by $\sigma$. There are three possibilities of internal settings
\bee{a9}
[\st{\bullet}{s},\mvec{J}],\ [\st{\bullet}{s},\sigma],\ [\sigma,\mvec{J}].
\ee
Both balances \R{a1} and the constitutive equation
\R{a4}$_1$ have to be taken into account for generating the entropy production \R{a8}. If
$[\sigma,\mvec{J}]$ is used for the internal setting, the resulting $\st{\bullet}{s}$ has to be a
total time differential. If not, the internal setting was not suitable and should be replaced and
repeated.\\ \\
{\bf 2. The procedure: Thermodynamical Verification}\\ 
The formal procedure presented in sect.1 is applicable to arbitrary materials. For short 
communication this scheme between \R{a1} and \R{a8} should be called {\em thermodynamical
verification of the entropy balance equation}. Now this is inspected in more detail.\\ \\
As an example we consider pure heat coduction with no power and material fluxes. Consequently,
the first internal setting is as \R{a4}$_2$ (resulting in \R{a5}$_2$)
\bee{a10}
\mvec{J}\ \seq\ (1/T)\mvec{q}\ \longrightarrow\ \nabla\cdot\mvec{J}\ =\ 
\mvec{q}\cdot\nabla(1/T)+(1/T)\nabla\cdot\mvec{q.} 
\ee
Taking the balances \R{a1} into account \R{a10}$_2$ results in \R{a6}
\byy{a11}
\sigma - \varrho\st{\bullet}{s}\ &\st{{2,1}}{=}&\ \mvec{q}\cdot\nabla(1/T)
+(1/T)(r-\varrho\st{\bullet}{u}),\\ \label{a12}\ 
\sigma &=&\ \mvec{q}\cdot\nabla(1/T)+(1/T)r
+\varrho\Big(\st{\bullet}{s}-(1/T)\st{\bullet}{u}\Big).
\eey
The second internal setting in \R{a12} is needed. Instead of \R{a7}$_1$, 
we now assume that $s$ beyond $u$ and ${\bf q}$ also depends of additional internal
variables $\mvec{\xi}$, that means
\bee{a15}
\st{\bullet}{s}\ \seq\ (1/T)\st{\bullet}{u}+\mvec{\alpha}\cdot\st{\bullet}{\mvec{q}}
+\mvec{\beta}\cdot\st{\bullet}{\mvec{\xi}}
\ee
is valid. Thus the entropy production \R{a12} becomes instead of \R{a7}$_2$
\bee{a14}
\sigma\ =\ \mvec{q}\cdot\nabla(1/T)+(1/T)r+\varrho\mvec{\alpha}\cdot\st{\bullet}{\mvec{q}}
+\varrho\mvec{\beta}\cdot\st{\bullet}{\mvec{\xi}}.
\ee
 
Now we check, if the thermodynamical verification was performed completely: There are two internal
settings in \R{a10}$_1$ and \R{a15} [$\st{\bullet}{s},\mvec{J}$]. The scheme of thermodynamical
verification is the same as in sect. 1. and 2., but the result depends on the special internal settings.
The balances \R{a1} are introduced in \R{a11}. A special constitutive equation such as \R{a4}$_1$
was up to now not taken into account and will be added later on.\\

The time differential of the entropy \R{a15} is in contrast to \R{a7}$_1$ defined on an enlarged set
of variables $\{u,\mvec{q},\mvec{\xi}\}$ which span a {\em state space} \C{Chris,MUAS1}, if the
variables are indepedent of each other. This is not evident because $u$ and $\mvec{q}$ occur jointly
in the energy balance. Clear is that $\st{\bullet}{\mvec{q}}$ and
\bee{a16}
\nabla\cdot\mvec{q}\ \st{1} {=}\ r-\varrho\st{\bullet}{u}\ \dashv\ \st{\bullet}{\mvec{q}}
\ee
are independent $\dashv$ of each other \C{Chris}. The opposite assumption that $u$ and
$\mvec{q}$ are connected with each other
\bee{a17}
u\ \dashv\ \hspace{-.5cm}\mvec{/}\hspace{.1cm}{\mvec{q}}\ \longrightarrow\ 
u\ =\ f(\mvec{q})\ \longrightarrow\ \st{\bullet}{u}\ =\ (df/d\mvec{q})\cdot\st{\bullet}{\mvec{q}}
\ee
results in \R{a17}$_3$. Inserting $\st{\bullet}{u}$ into \R{a16} yields an expression which depends
on $\st{\bullet}{\mvec{q}}$
\bee{a17a}
r-\varrho (df/d\mvec{q})\cdot
\st{\bullet}{\mvec{q}}\ \dashv\ \hspace{-.5cm}\mvec{/}\hspace{.1cm}\st{\bullet}{\mvec{q}},
\ee
a statement which is in contradiction to \R{a16}. Consequently, the assumption of contradiction
\R{a17}$_1$ that $u$ and $\mvec{q}$ depend on each other is wrong: they are independet of each
other
\bee{a18}
u\ \dashv\ \mvec{q}.
\ee

Consequently, $\{u,\mvec{q},\mvec{\xi}\}$ can be used as a state space, and the time differential
of the entropy $s(u,\mvec{q},\mvec{\xi})$ \R{a15} is a total one
\bee{a19}
(\partial s/\partial u)\ =\ (1/T),\ \ (\partial s/\partial\mvec{q})\ =\ \mvec{\alpha},\ \  
(\partial s/\partial\mvec{\xi})\ =\ \mvec{\beta}.
\ee
Now two additional constitutive settings are introduced
\bee{d20}
\mvec{\alpha}\ \st{c}{=}\ \mbox{{\bf c}onst},\ \ \mvec{\beta}\ \st{c}{=}\ \mbox{{\bf k}onst}.
\ee
Taking \R{a19} into account, the second mixed differentials result in
\byy{a20}
(\partial /\partial\mvec{q})(1/T) &=& (\partial /\partial u)\mvec{\alpha}\ =\ \mvec{0},
\\ \label{a21}
(\partial /\partial\mvec{\xi})(1/T) &=& (\partial /\partial u)\mvec{\beta}\ =\ \mvec{0},
\\ \label{a22}
(\partial /\partial\mvec{\xi})\mvec{\alpha} &=& (\partial /\partial \mvec{q})\mvec{\beta}\ 
=\ \mvec{0}.
\eey

The zeros in \R{a20} to \R{a22} are generated by the constitutive settings \R{d20}.
From \R{a20}$_1$ and \R{a21}$_1$ follows
\bee{d24}
T\ =\ F(u)\ \longrightarrow\ \st{\bullet}{T}\ =\ \st{\bullet}{u}(\partial /\partial u)F\ 
\longrightarrow\ \frac{\st{\bullet}{T}}{T(\partial /\partial u)F}\ =\ \st{\bullet}{u}/T,
\ee
the possibility to change the state space
\bee{d25}
\{u,\mvec{q},\mvec{\xi}\}\ \longrightarrow\ \{T,\mvec{q},\mvec{\xi}\}.
\ee
The time differential of the entropy \R{a15} becomes
\bee{d26}
\st{\bullet}{s}\ =\ \frac{\st{\bullet}{T}}{T(\partial /\partial u)F}+\mvec{\alpha}\cdot\st{\bullet}{\mvec{q}}
+\mvec{\beta}\cdot\st{\bullet}{\mvec{\xi}}.
\ee
This trivial example shows that state spaces represent an essential tool of material description.\\ \\
{\bf 3. A special [$\st{\bullet}{s},\mvec{J}$]-verification using an extra entropy flux}\\
The first internal setting introduces the {\em extra entropy flux} $\mvec{K}$ generalizing
\R{a4}$_2$ and \R{a10}$_1$. Taking \R{a6} into consideration, one  obtains \R{a26} 
\byy{a24}
\mvec{J} &\seq& (1/T)\mvec{q}+\mvec{K},
\\ \label{a25}
\nabla\cdot\mvec{J} &=& \mvec{q}\cdot\nabla(1/T)+(1/T)\nabla\cdot\mvec{q}
+ \nabla\cdot\mvec{K}\ \st{2}{=}\\ \label{a26}
&\st{2}{=}& \sigma-\varrho\st{\bullet}{s}\ \st{1}{=}\ 
\mvec{q}\cdot\nabla(1/T)+(1/T)(r-\varrho\st{\bullet}{u})
+ \nabla\cdot\mvec{K}.
\eey

Presupposing that the extra entropy flux has two parts, one which is not parallel to the heat flux and
the other which is parallel to the time differential of the heat flux, we obtain the constitutive
setting
\byy{a27}
\mvec{K} &\st{c}{=}& \mathbb{Q}\cdot\mvec{q}+a\st{\bullet}{\mvec{q}},
\\ \label{28}
\nabla\cdot\mvec{K} &=&
\nabla\cdot(\mathbb{Q}\cdot\mvec{q})+\nabla\cdot (a\st{\bullet}{\mvec{q}})\ =\\ \label{a29}
&=&\mvec{q}\nabla:\mathbb{Q}^{\top}+\mathbb{Q}:\nabla\mvec{q}\ +
\st{\bullet}{\mvec{q}}\cdot\nabla a + a\nabla\cdot\st{\bullet}{\mvec{q}}.
\eey
\footnote{$\mathbb{A}:\mathbb{B}\ \rightarrow\ A_{kl}B_{kl}$}
Now the second internal setting follows from \R{a26} by inserting \R{a29}
\bee{a30}
-\varrho\st{\bullet}{s}\ \seq\ -\varrho(1/T)\st{\bullet}{u}+
\st{\bullet}{\mvec{q}}\cdot\nabla\sf{a}\ \longrightarrow\ \label{d33}\ 
\st{\bullet}{s}\ \seq (1/T)\st{\bullet}{u}-(1/\varrho)
\st{\bullet}{\mvec{q}}\cdot\nabla\sf{a},
\ee
establishing the state space $\{u,\mvec{q}\}$\footnote{remember: u and $\mvec{q}$ are
independent of each other}:
\bee{d34}
a\ =\ a(u,\mvec{q}),\ \ \ \mathbb{Q}\ =\ \mathbb{Q}(u,\mvec{q}).
\ee
Taking \R{a29} and \R{a30}$_1$ into account, \R{a26} results in the entropy production
\bee{a31}
\sigma\ =\ \Big(\nabla(1/T)+\nabla\cdot\mathbb{Q}\Big)\cdot\mvec{q}
+(1/T)r+\mathbb{Q}:\nabla\mvec{q}\ + a\nabla\cdot\st{\bullet}{\mvec{q}}
\ee
which has an interesting shape in comparison to \R{a7}: the gradient of the
reciprocal temperature is supplemented by the divergence of that tensor
$\mathbb{Q}$ which destroys the parallelism of the extra entropy flux with the heat flux.

Additional constitutive settings
\bee{d36}
a\ \st{c}{=}\ const,\ \ \ \mathbb{Q}\ \st{c}{=}\ \mathbb{K}onst,
\ee
change the state space, the time differential of the entropy \R{d33}$_2$ and the entropy
production \R{a31}
\bee{d37}
\st{\bullet}{s}\ \seq\ (1/T)\st{\bullet}{u},\ \ \ \sigma\ =\ \nabla(1/T)\cdot\mvec{q}
+(1/T)r+\mathbb{Q}:\nabla\mvec{q}\ + a\nabla\cdot\st{\bullet}{\mvec{q}}.
\ee
If aditionally $a$ and $\mathbb{Q}$ vanish, Fourier heat conduction of sect. 1 emerges.\\

\noindent
{\bf Acknowledgment}\\
\noindent
My warm thanks to Prof. Dr. Karl Heinz Hoffmann for helpful discussions on the utilization of
thermodynamical verifications and for reviewing a former version of this short note.

\end{document}